\documentclass[12pt,a4 paper]{article}
\usepackage{graphics}
\usepackage{epsfig}
\begin{document}
\title{Comments on Fractality of Proton at Small $x$}
\author{\it{D.K.Choudhury and  Rupjyoti Gogoi}\footnote{Corresponding author: rupjyoti2000@yahoo.com}\\\it{Department of Physics, Gauhati University, Guwahati-781 014}}
\maketitle

\begin{abstract}
Using the concept of self similarity in the structure of the proton at small $x$, we comment on possibility of a single positive fractal dimension of proton in analogy with classical monofractals. Plausible dynamics and physical interpretation of fractal dimension are also discussed.
\end{abstract}
\newpage
Self similarity is a familiar property in nature [1, 2]. Many of the seemingly irregular shapes of nature have hidden self similarity in them. It is not the usual symmetry with respect to rotation or translation, but symmetry with respect to scale or size: a small part of a system is self similar to the entire system. Such a system is defined through its self similar dimension, which is in general fraction, hence called fractal dimension. Classical fractals discussed in standard references [1, 2] are Cantor dust, Koch curve and Sierpinski gasket whose fractional dimensions are 0.63, 1.26 and 1.585 respectively, which lie between Eucledian point and surfaces.

Notion of self similarity and fractal dimensions are being used in the phase spaces of hadron multiparticle production processes since nineteen eighties [3-7]. However these ideas did not attract much attention in contemporary physics of deep inelastic lepton hadron scattering, due to its obvious lack of applicability. Only recently [8], Lastovicka has developed relevant formalism and applied it to deep inelastic electron-proton scattering at small $x$ and proposed a functional form of the structure function $F_{2}(x,Q^{2})$. Specifically a description of $F_{2}(x,Q^{2})$ reflecting self similarity is proposed with a few parameters which are fitted to recent HERA data [9, 10]. The specific parameterization is claimed to provide an excellent description of the data which covers a  region of four momentum transferred squared $0.045 \leq Q^{2}\leq 150GeV^{2}$ and of Bjorken $x$, $6.2\times 10^{-7}\leq x\leq0.2$.
 More recently, it was observed [11-13] that the positivity of fractal dimensions prohibits some of the fitted parameters of the structure function of reference [8]. Specifically out of the fractal dimensions $D_{1}$, $D_{2}$ and $D_{3}$, one is negative ($D_{3} \approx -1.3$). However the positivity of fractal dimension forbids such negative value. In order to avoid such possibility, it is suggested that the proton is described by the single self similarity dimension $D$. This then facilitates one to compare the self similarity nature of the proton at small $x$ with the classical monofractals which is the aim of the present note.

Under the hypothesis of self similarity of the proton structure at small $x$, Lastovicka [8] obtained the following form of the structure function $F_{2}(x,Q^{2})$,

\begin{equation}
F_{2}=\frac{\left(\exp{D_{0}}\right){Q_{0}}^{2}x^{-D_{2}+1}}{1+D_{3}+D_{1}\log{\frac{1}{x}}}\left(x^{{-D_{1}}\log{\left(1+\frac{Q^{2}}{{Q_{0}}^{2}}\right)}}{\left(1+\frac{Q^{2}}{{Q_{0}}^{2}}\right)}^{D_{3}+1}-1\right).
\end{equation}
by using the following form of un-integrated quark density $f_{i}(x,Q^{2})$ of $i$-th quark flavor:
\begin{equation}
\log{f_{i}(x,Q^{2})}=D_{1}\log{\frac{1}{x}}\log{\left(1+\frac{Q^{2}}{{Q_{0}}^{2}}\right)}+D_{2}\log{\frac{1}{x}}+D_{3}\log{\left(1+\frac{Q^{2}}{{Q_{0}}^{2}}\right)}+{D_{0}}^{i}.
\end{equation}
In (1) and (2), $D_{1}$ is the dimensional correlation relating the two magnification factors $\frac{1}{x}$ and $\left(1+\frac{Q^{2}}{{Q_{0}}^{2}}\right)$, while $D_{2}$ and $D_{3}$ are the self similarity dimensions associated with them
, ${D_{0}}^{i}$ being the normalisation constant. Since the magnification factors should be positive, non-zero and dimensionless, a choice $1+\frac{Q^{2}}{{Q_{0}}^{2}}$, rather than $Q^{2}$ has been made, while ${Q_{0}^{2}}$ is arbitrary small virtuality, $Q^{2}>{Q_{0}}^{2}$. Explicit confrontation with HERA data [9, 10] yields,
\begin{eqnarray}
D_{0}&=&0.339\pm0.145\nonumber\\
D_{1}&=&0.073\pm0.001\nonumber\\
D_{2}&=&1.013\pm0.01\nonumber\\
D_{3}&=&-1.287\pm0.01\nonumber\\
{Q_{0}}^{2}&=&0.062\pm0.01GeV^{2}.
\end{eqnarray}
As the self similarity dimensions of fractals are positive [1, 2], by its definitions one expects $D_{1}\geq0$, $D_{2}\geq0$, $D_{3}\geq0$, a feature absent in the empirical fit of [8] as far as $D_{3}$ is concerned. In analogy with other classic fractals [1, 2] we therefore assume that proton at small $x$ is a monofractal with just one single fractal dimension, so that 

\begin{equation}
D_{1}=D_{2}=D_{3}=D.
\end{equation}

Under such a hypothesis, equation (1) is rewritten as,

\begin{equation}
F_{2}=\frac{\left(\exp{D_{0}}\right){Q_{0}}^{2}x^{-D+1}}{1+D+D\log{\frac{1}{x}}}\left(x^{{-D}\log{\left(1+\frac{Q^{2}}{{Q_{0}}^{2}}\right)}}{\left(1+\frac{Q^{2}}{{Q_{0}}^{2}}\right)}^{D+1}-1\right).
\end{equation}

Alternately, monofractality is attainable also for $D_{1}=0$ (zero dimensional correlation) and $D_{2}=D_{3}=D$, so that equation (1) takes the alternate form,
\begin{equation}
  F_{2}(x,Q^{2})=\frac{\left(\exp{D_{0}}\right){Q_{0}}^{2}x^{-D+1}}{1+D}\left(\left(1+\frac{Q^{2}}{{Q_{0}}^{2}}\right)^{D+1}-1\right).
\end{equation}

In figure (1), we plot $F_{2}(x,Q^{2})$ versus $x$ in bins of $Q^{2}$ as measured by low $Q^{2}$ data of ZEUS [10] using Equation(5).
Results of the fit yields,
\begin{eqnarray}
D_{0}&=&-1.692\pm0.14\nonumber\\
D&=&0.653\pm0.029\nonumber\\
{Q_{0}}^{2}&=&0.0449\pm0.0003GeV^{2}.
\end{eqnarray}
However this fit(Equation 7)  can not be extrapolated to higher $Q^{2}$ range of H1 [9]. Even for $Q^{2}>0.4GeV^{2}$ of ZEUS [10] data, $\chi^{2}$ becomes large.\\
\begin{center}
Table 1
\end{center}
\begin{center}
\begin{tabular}{|c|c|c|}\hline
fit&$\chi^{2}$& $\chi^{2}/dof$ \\\hline
equation7($0.045\leq Q^{2}\leq0.4GeV^{2}$)&78.504 &1.402 \\\hline
equation8($0.045\leq Q^{2}\leq12GeV^{2}$)&83.156 &0.621\\\hline
\end{tabular}
\end{center}

In figure (2), we show the similar analysis using equation (6) for ZEUS data [10]. Results of the fit yields,
\begin{eqnarray}
D_{0}&=&-2.713\pm0.231\nonumber\\
D_{1}&=&0\nonumber\\
D&=&1.107\pm0.008\nonumber\\
{Q_{0}}^{2}&=&0.045\pm0.00012GeV^{2}.
\end{eqnarray}
This fit (Equation 8) can be extrapolated to higher $Q^{2}$ range of H1 [10] upto $Q^{2}=12GeV^{2}$. The $\chi^{2}$ for equation (7) and (8) are recorded in table 1.

Our analysis thus indicates that only in the limited $x-Q^{2}$ range ($Q^{2}\leq12GeV^{2}$), the notion of monofractality of proton holds. In that range, dimensional correlation ($D_{1}$) vanishes and the proton possesses fractality ($D\approx 1.107$) close to Koch curve ($D \approx 1.26$). Description of $F_{2}(x,Q^{2})$ in the entire small $x$ range in terms of monofractal will result in a continuous $x$, $Q^{2}$ dependent fractal dimension [13] which is a considerable extension of parameter space and contrary to the usual notion of fractal. From dynamical point of view, breakdown of monofractality conjecture presumbly implies existence of long range interaction among the small $x$ gluons. It is well known that [14] if there are no long range interactions , one expects that all fractal dimensions are equal to one another; i.e. the system is a single fractal as demonstrated explicitly by the Ising model [15] and Feynman Wilson fluid [16].

It is also instructive at this stage to ascertain the physical interpretation of fractal dimension of proton, since the notion is rather recent in literature. As is well known [17], the fractal dimension measures the way, in which distribution of points fill a geometric space on the average. If the distribution is highly inhomogeneous, the set of points have a distribution of fractal dimensions leading to multifractality. Extending the notion to the $x-Q^{2}$ plane of the unintegrated quark density, fractal dimension tells how densely small $x$ partons fill the proton in self similar way on the average. In the special case of $D\approx D_{2}>>D_{1}$,$D_{3}$, unintegrated quark density takes the simple form,
\begin{equation}
f(x,Q^{2})\approx\left(\frac{1}{x}\right)^{D}.
\end{equation}
and fractal dimension is essentially close to $x$-slope [18] or pomeron intercept [19-21].

To conclude, let us also comment on the negative fractal dimension $D_{3}$ of reference [8]. As noted earlier, fractal dimensions being positive, Lastovicka's work [8] as such cannot be regarded strictly as pure fractal analysis. It is  a good description of data in terms of five parameters without implication on fractals. One plausible way of preserving positivity of fractal dimensions but still giving a good description of HERA data is through suitable modification of the magnification factors as occurred in the original formalism. Such a work is currently in progress.

\section*{Acknowledgement}
We thank Dr. T. Lastovicka for useful correspondences.

\begin{figure}[t]
\begin{center}
\includegraphics[height=6cm,width=4cm]{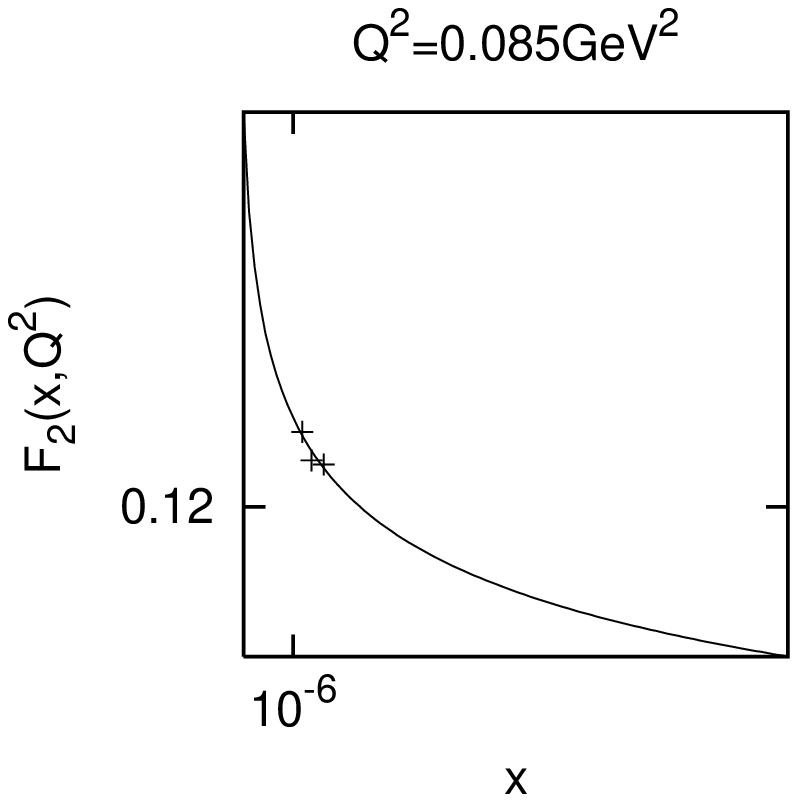}
\includegraphics[height=6cm,width=4cm]{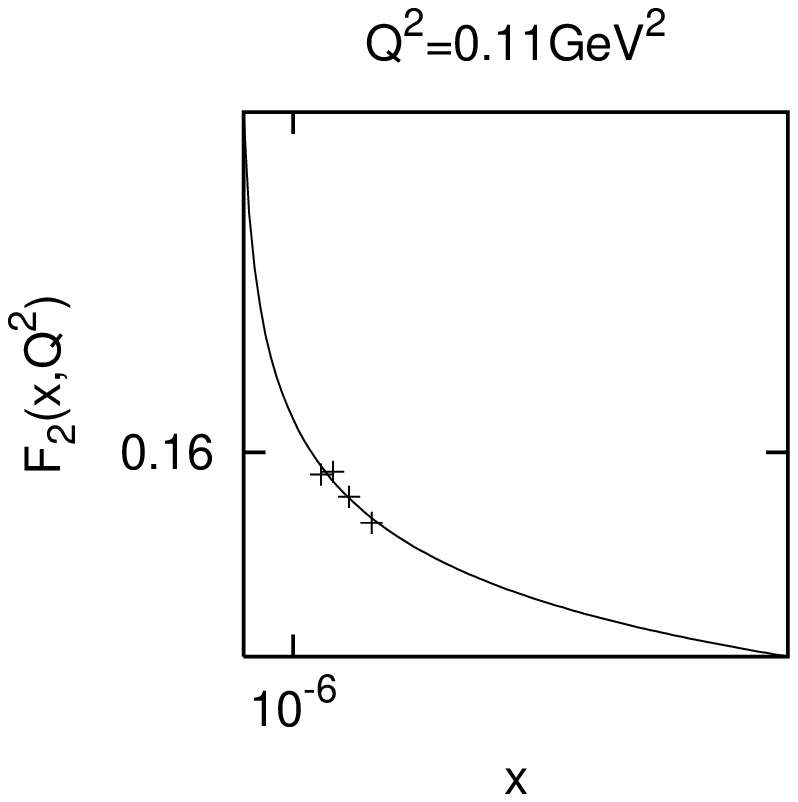}
\includegraphics[height=6cm,width=4cm]{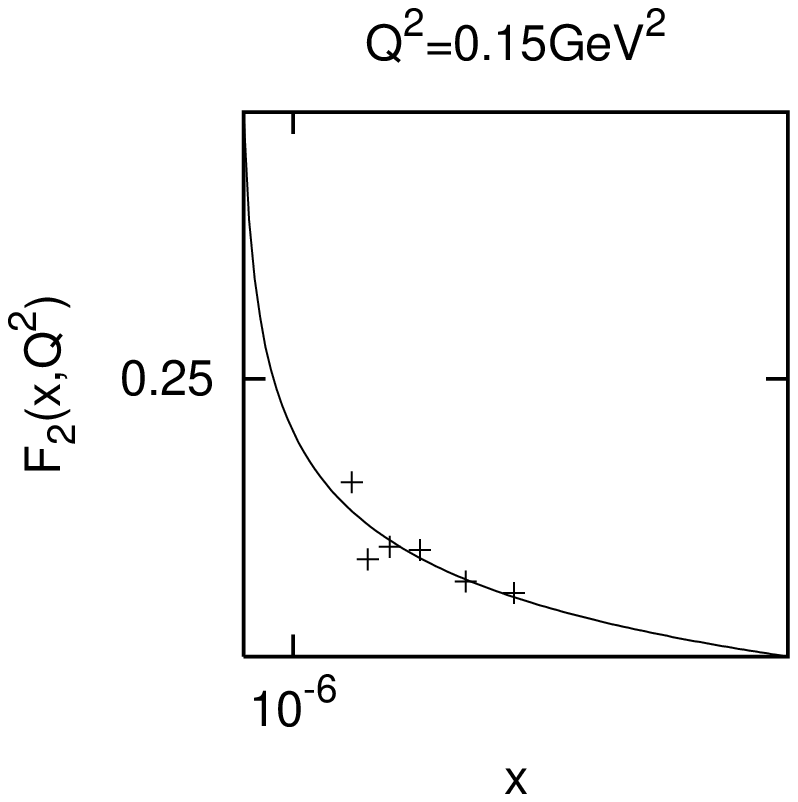}
\includegraphics[height=6cm,width=4cm]{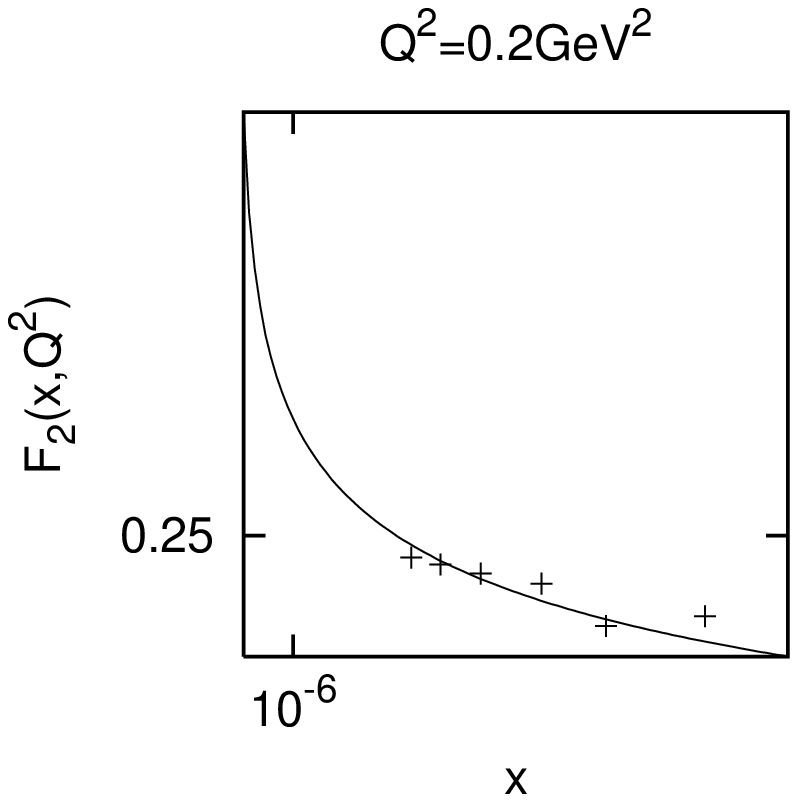}
\includegraphics[height=6cm,width=4cm]{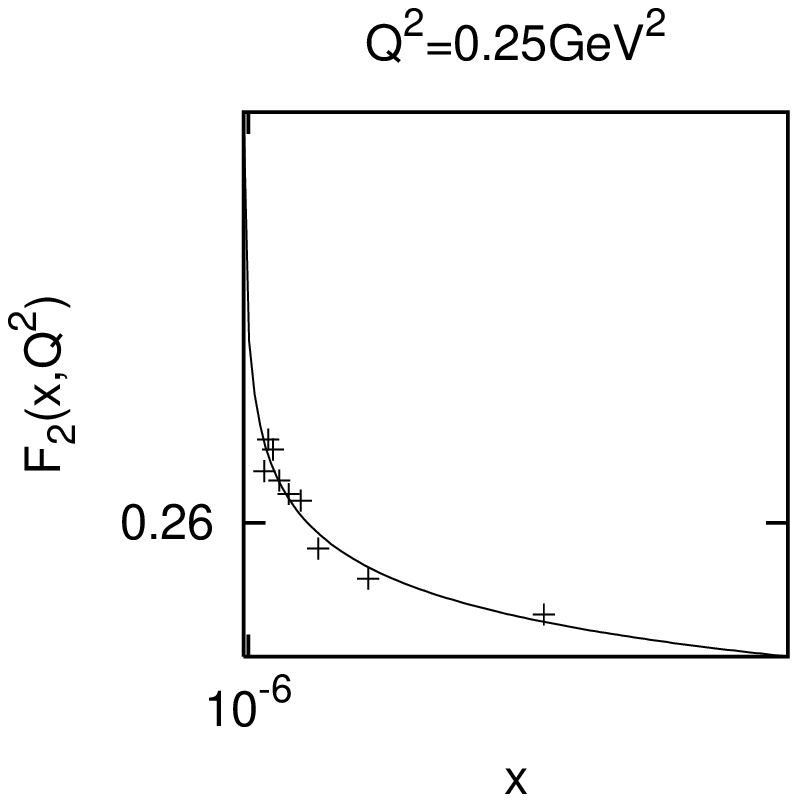}
\includegraphics[height=6cm,width=4cm]{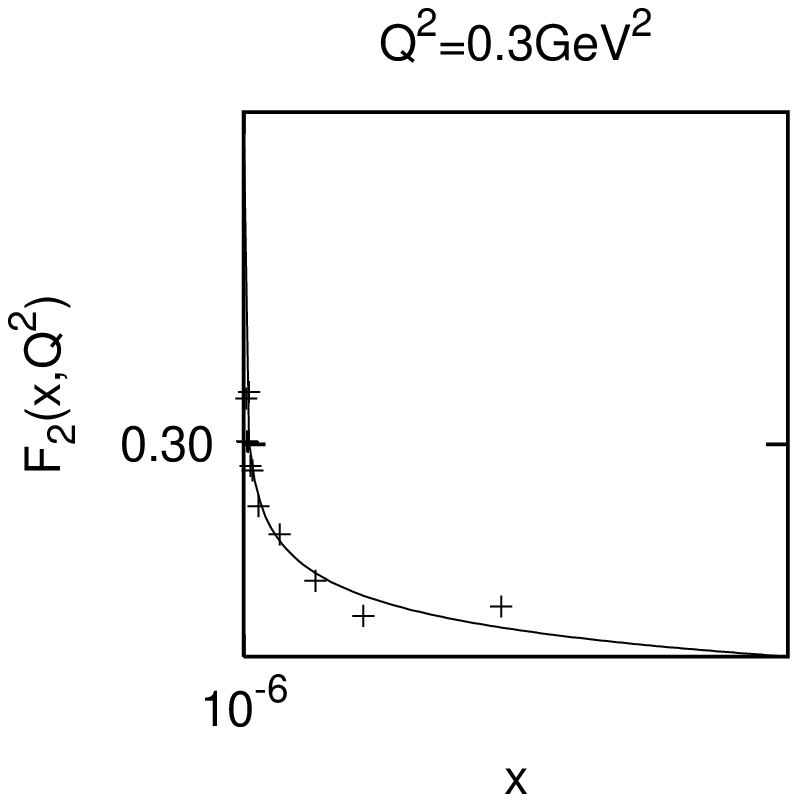}
\includegraphics[height=6cm,width=4cm]{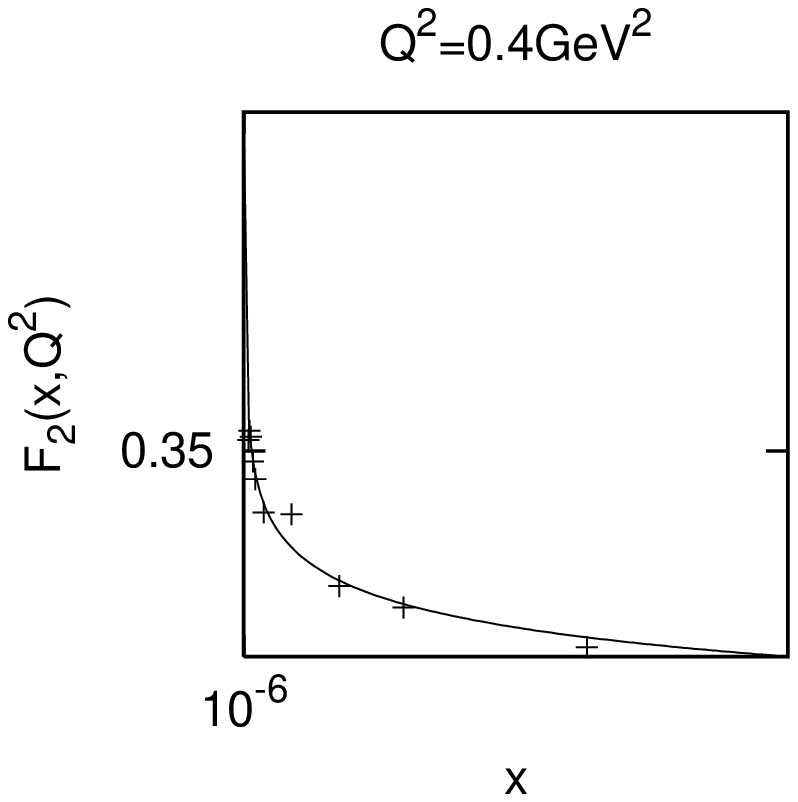}
\includegraphics[height=6cm,width=4cm]{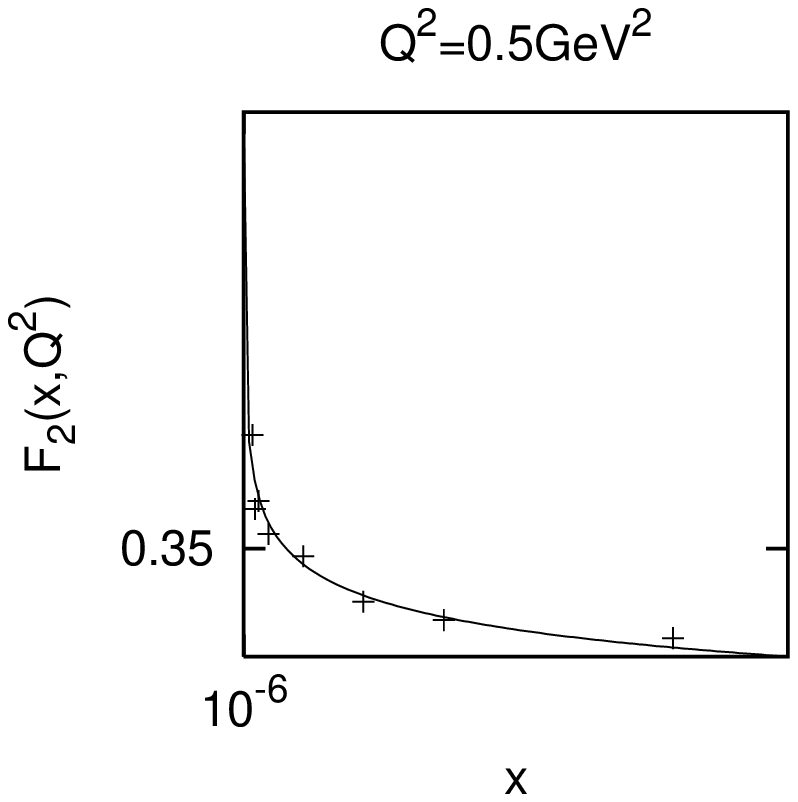}
\includegraphics[height=6cm,width=4cm]{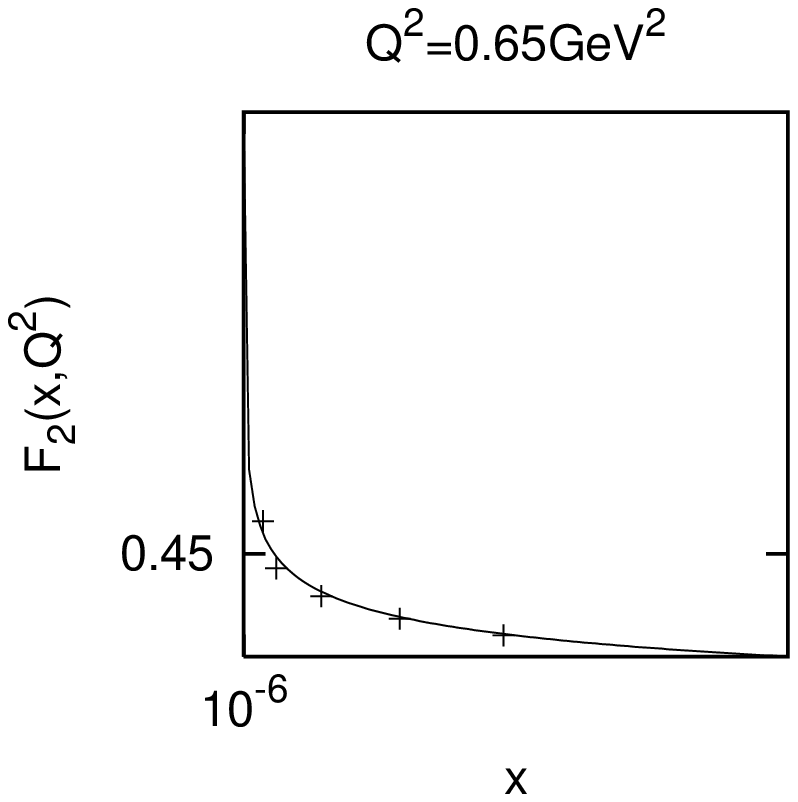}
\end{center}
\caption{$F_{2}(x,Q^{2})$ versus $x$ in bins of $Q^{2}$ with $D_{1}\neq0$ (Equation 5)}
\end{figure}
\begin{figure}[t]
\begin{center}
\includegraphics[height=6cm,width=4cm]{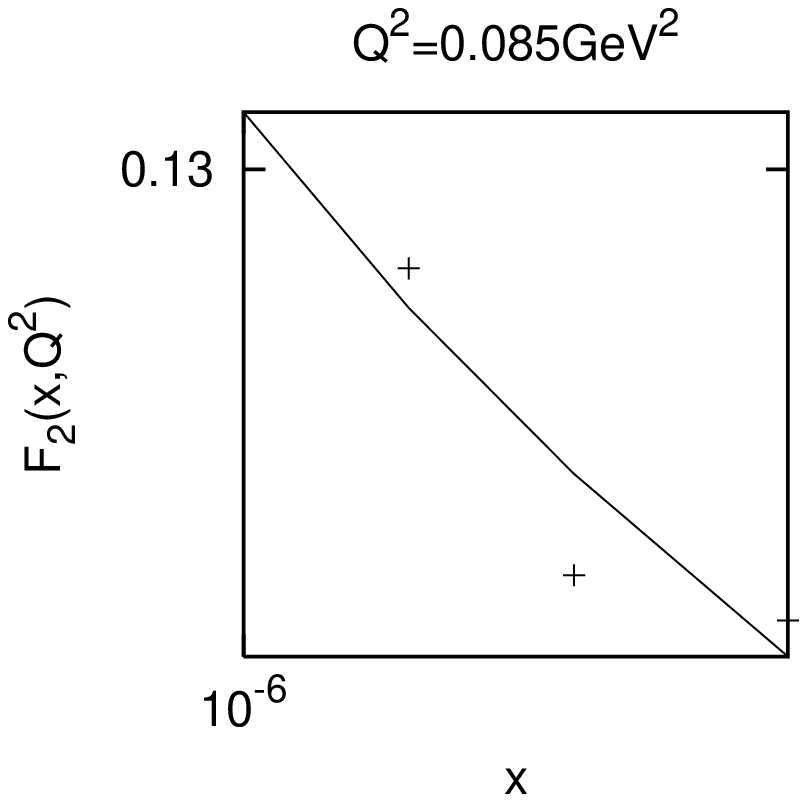}
\includegraphics[height=6cm,width=4cm]{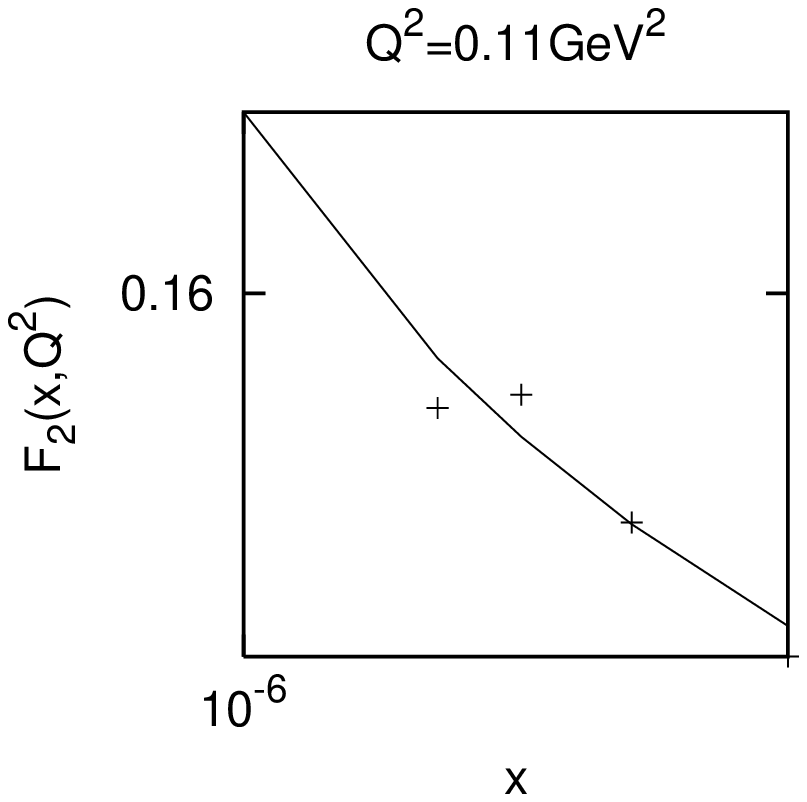}
\includegraphics[height=6cm,width=4cm]{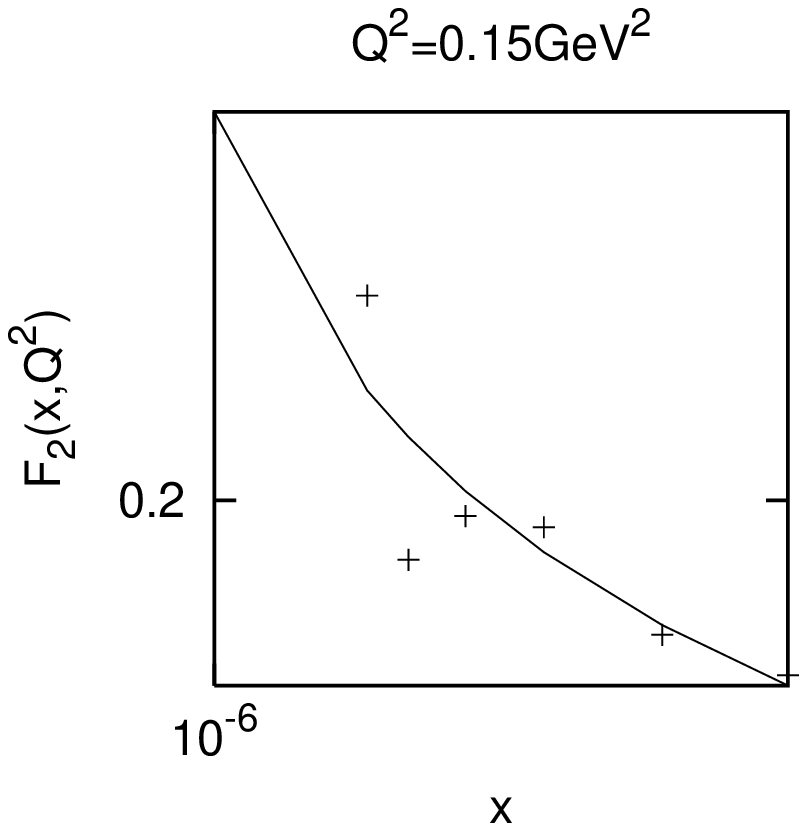}
\includegraphics[height=6cm,width=4cm]{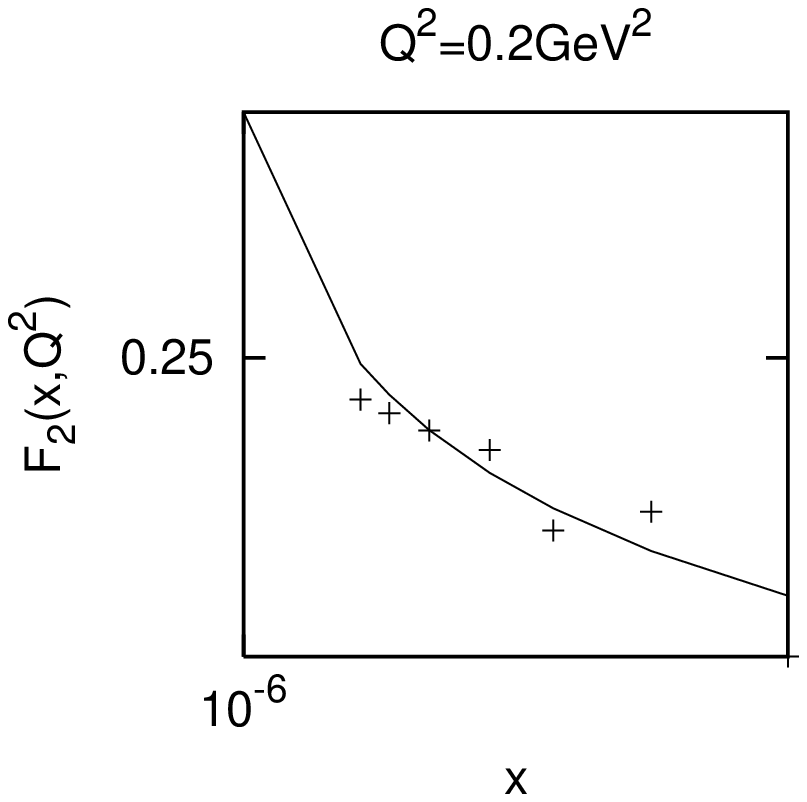}
\includegraphics[height=6cm,width=4cm]{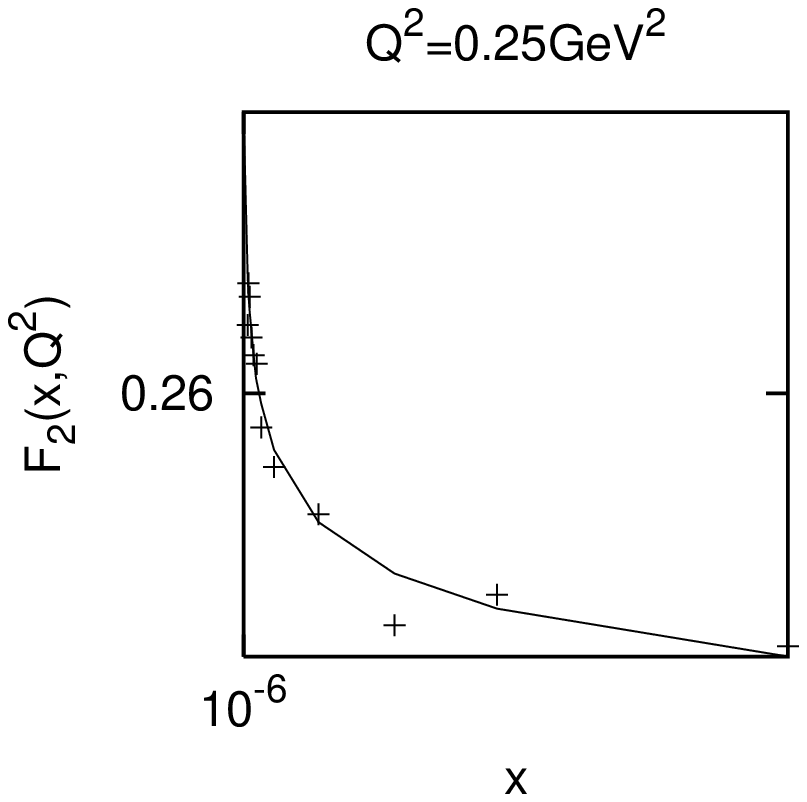}
\includegraphics[height=6cm,width=4cm]{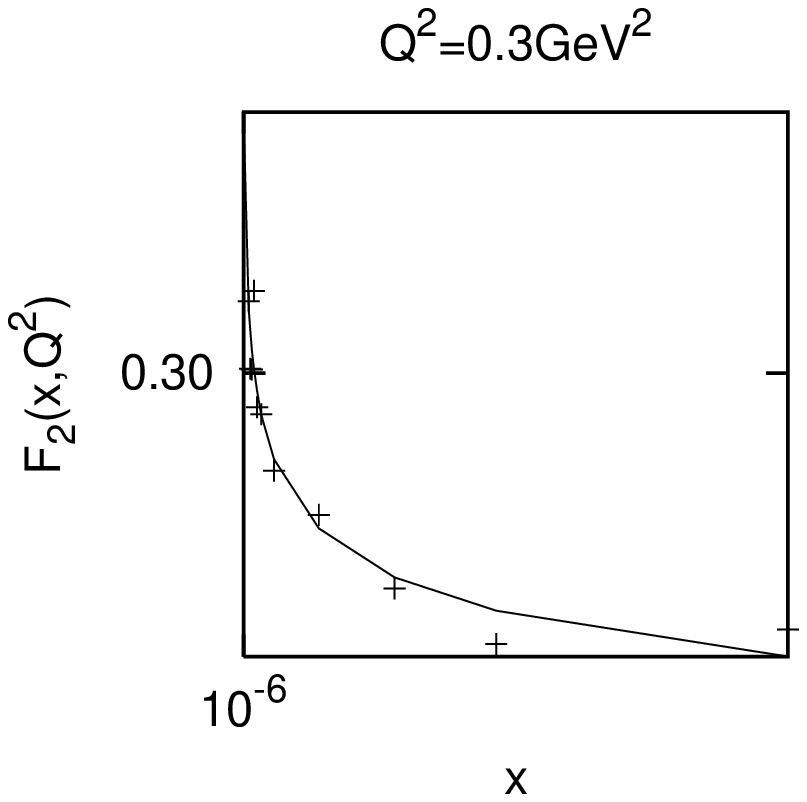}
\includegraphics[height=6cm,width=4cm]{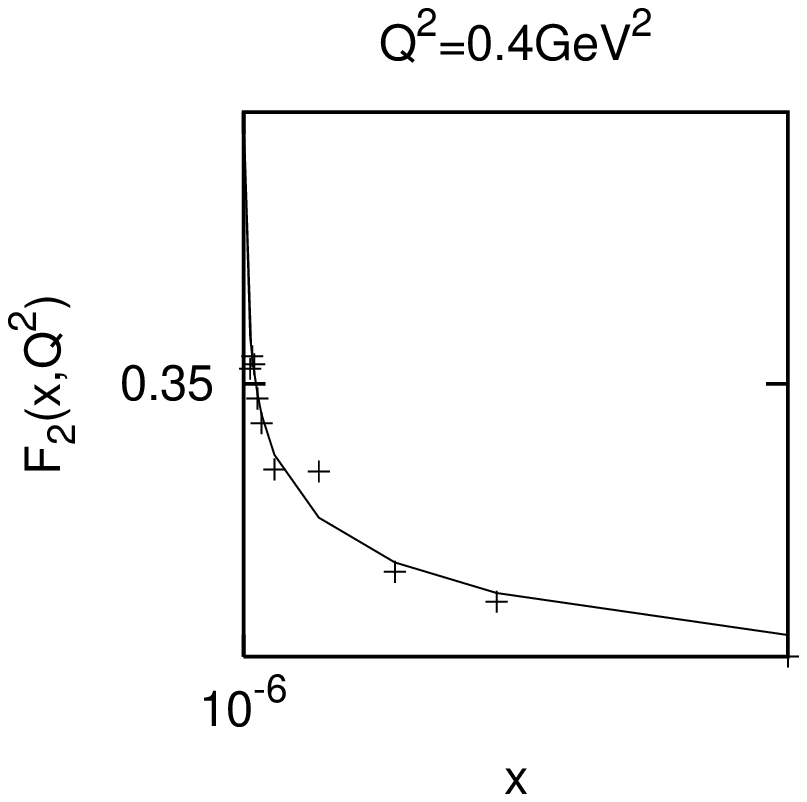}
\includegraphics[height=6cm,width=4cm]{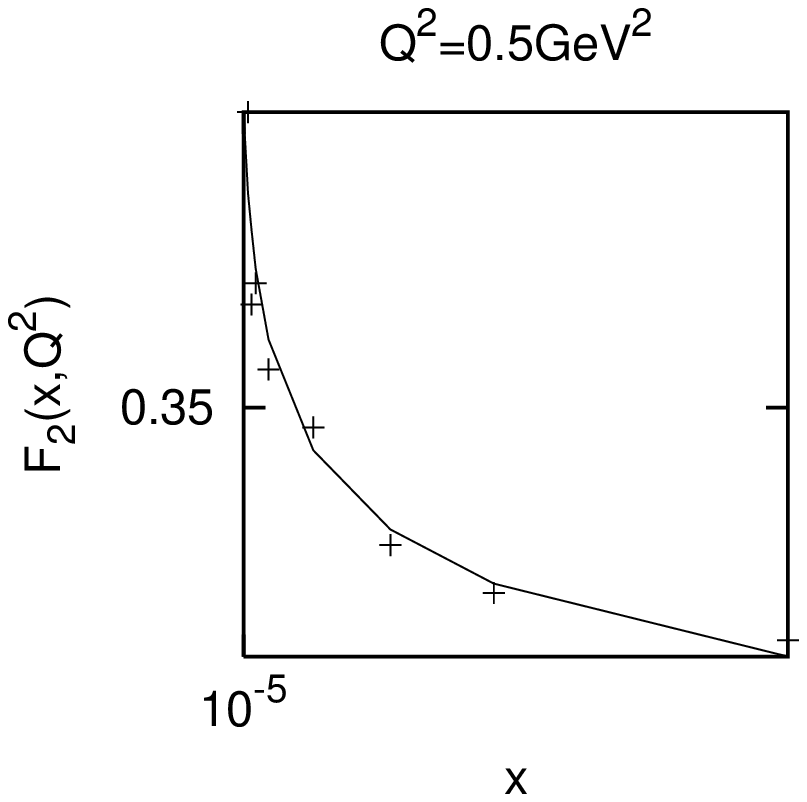}
\includegraphics[height=6cm,width=4cm]{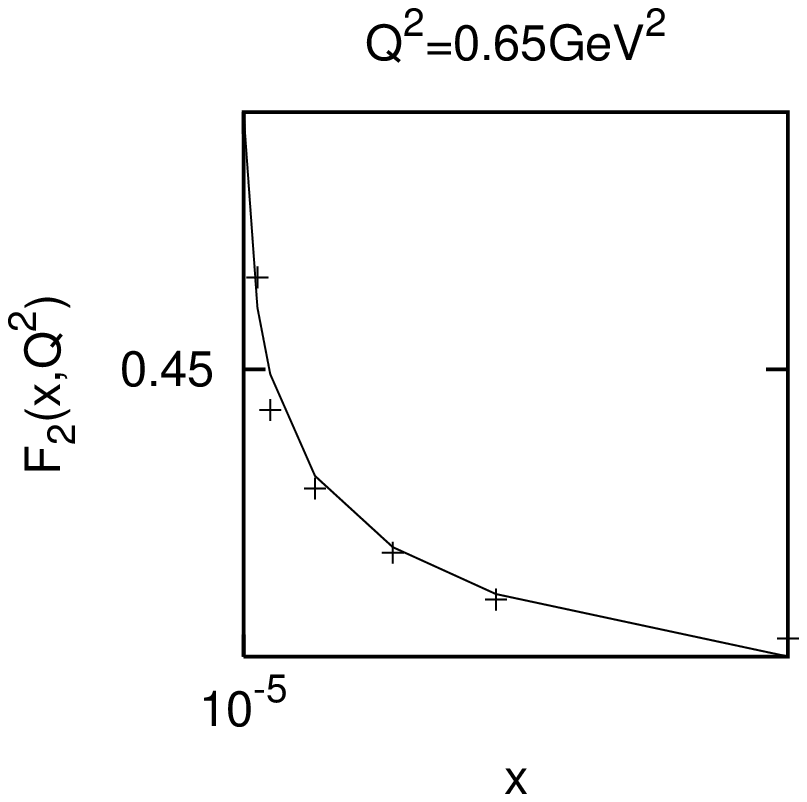}
\end{center}
\caption{$F_{2}(x,Q^{2})$ versus $x$ in bins of $Q^{2}$ with $D_{1}=0$ (Equation 6)}
\end{figure}

\end{document}